\documentclass[10pt,twocolumn,letterpaper]{article}

\usepackage{iccv}
\usepackage{times}
\usepackage{epsfig}
\usepackage{graphicx}
\usepackage{amsmath}
\usepackage{amssymb}

\usepackage{multirow}
\usepackage{setspace}
\usepackage[table,xcdraw]{xcolor}
\usepackage{subfigure}
\usepackage[pagebackref=true,breaklinks=true,colorlinks,bookmarks=false]{hyperref}
\usepackage[accsupp]{axessibility}  

\usepackage[breaklinks=true,bookmarks=false]{hyperref}
\usepackage{pdfpages}

\iccvfinalcopy 

\def\iccvPaperID{6924} 

\ificcvfinal\fi

\begin{document}

\title{A New Journey from SDRTV to HDRTV}

\author{
 Xiangyu Chen\textsuperscript{\rm 1\rm *} \space \space 
 Zhengwen Zhang\textsuperscript{\rm 1\rm *} \space \space
 Jimmy S. Ren\textsuperscript{\rm 2,3} \space \space
 Lynhoo Tian\textsuperscript{\rm 2} \space \space
 Yu Qiao\textsuperscript{\rm 1,4}   \space \space 
 Chao Dong\textsuperscript{\rm 1\dag}\\
 \textsuperscript{\rm 1}ShenZhen Key Lab of Computer Vision and Pattern Recognition, SIAT-SenseTime Joint Lab,\\Shenzhen Institute of Advanced Technology, Chinese Academy of Sciences
 \textsuperscript{\rm 2}SenseTime Research\\ 
 \textsuperscript{\rm 3}Qing Yuan Research Institute, Shanghai Jiao Tong University \space \space
 \textsuperscript{\rm 4}Shanghai AI Laboratory, Shanghai\\
 {\tt\small\{chxy95, zhengwen.zhang02, jimmy.sj.ren, lynhoo.tian\}@gmail.com}\space \space
 {\tt\small\{yu.qiao, chao.dong\}@siat.ac.cn} 
}

\maketitle
\ificcvfinal\thispagestyle{empty}\fi

\renewcommand{\thefootnote}{\fnsymbol{footnote}}
\footnotetext[1]{indicates contribute equally. $^\dag$Corresponding author.}
\renewcommand{\thefootnote}{1}

\begin{abstract}
Nowadays modern displays are capable to render video content with high dynamic range (HDR) and wide color gamut (WCG). However, most available resources are still in standard dynamic range (SDR). Therefore, there is an urgent demand to transform existing SDR-TV contents into their HDR-TV versions. In this paper, we conduct an analysis of SDRTV-to-HDRTV task by modeling the formation of SDRTV/HDRTV content. Base on the analysis, we propose a three-step solution pipeline including adaptive global color mapping, local enhancement and highlight generation. Moreover, the above analysis inspires us to present a lightweight network that utilizes global statistics as guidance to conduct image-adaptive color mapping. In addition, we construct a dataset using HDR videos in HDR10 standard, named HDRTV1K, and select five metrics to evaluate the results of SDRTV-to-HDRTV algorithms. Furthermore, our final results achieve state-of-the-art performance in quantitative comparisons and visual quality. The code and dataset are available at \url{https://github.com/chxy95/HDRTVNet}.

\end{abstract}

\section{Introduction}
The resolution of television (TV) content has increased from standard definition (SD) to high definition (HD) and most recently to ultra-high definition (UHD). High dynamic range (HDR) is one of the biggest features of the latest TV generation. HDRTV\footnote{We add a suffix TV after HDR/SDR to indicate content in HDR-TV/SDR-TV format and standard.} content has much wider color space and higher dynamic range than SDR content, and HDRTV standard allows us to create images and videos that are closer to what we see in real life. While HDR display devices have become prevalent in daily life, most accessible resources are still in SDR format. Therefore, we need algorithms to convert SDRTV content to their HDRTV version. The task, denoted as SDRTV-to-HDRTV, is of great practical value, but received relatively less attention in the research community. The reason is mainly two-fold. First, existing HDRTV standards (e.g., HDR10 and HLG) have not been well defined until recent years. Second, there is a lack of large-scale datasets for training and testing. This work aims at promoting the development of this emerging area, by presenting the analysis of this problem, basic methods, evaluation metrics and a new dataset. 

SDRTV-to-HDRTV is a highly ill-posed problem. In actual production, contents of SDRTV and HDRTV are derived from the same Raw file but are processed under different standards. Thus, they have different dynamic ranges, color gamuts and bit-depths. To some extent, SDRTV-to-HDRTV is similar to image-to-image translation such as Pixel2Pixel \cite{isola2017image} and CycleGAN \cite{zhu2017unpaired}. On the contrary, the task of LDR-to-HDR, which is similar in terms of name, has completely different connotations. LDR-to-HDR methods \cite{kovaleski2014high, masia2009evaluation, huo2014physiological, liu2020single, eilertsen2017hdr} aim to predict the HDR scene luminance in the linear domain, which is closer to Raw file in essence. SDRTV-to-HDRTV has recently been touched in Deep SR-ITM \cite{kim2019deep} and JSI-GAN \cite{kim2020jsi}, which try to solve the problem of joint super-resolution and SDRTV-to-HDRTV. Although the above-mentioned works are all related to SDRTV-to-HDRTV, they are not dedicated to this problem. We will detail these comments in Sec. \ref{Preliminary} and Sec. \ref{Analysis}.

This paper aims to address SDRTV-to-HDRTV based on deep understanding of this problem. We first provide a simplified formation pipeline for SDRTV/HDRTV content, which consists of tone mapping, gamut mapping, transfer function and quantization, as in Fig. \ref{Figure 1 (a)}. Based on the formation pipeline, we propose a solution pipeline, including adaptive global color mapping (AGCM), local enhancement (LE) and highlight generation (HG). For AGCM, we propose a novel color condition block to extract global image prior and adapt different images. With only $1\times1$ filters, the network achieves the best performance with less parameters compared with other photo retouching methods such as CSRNet \cite{he2020conditional}, HDRNet \cite{gharbi2017deep} and Ada-3DLUT \cite{zeng2020learning}. Besides, we adopt a commonly used ResNet-based network and a GAN-based model for LE and HG, respectively.

To promote the progress of this new research area, we collect a new large-scale dataset, named HDRTV1K. We also select five evaluation metrics -- PSNR, SSIM, SR-SIM \cite{zhang2012sr}, $\Delta E_{ITP}$ \cite{ITP} and HDR-VDP3 \cite{mantiuk2011hdr}, to evaluate the mapping accuracy, structural similarity (SSIM and SR-SIM), color difference and visual quality, respectively.

Our contributions are four-fold: (1) We conduct a detailed analysis for SDRTV-to-HDRTV task by modeling the formation of SDRTV/HDRTV content. (2) We propose a three-step SDRTV-to-HDRTV solution pipeline and a corresponding method, which performs best in quantitative and qualitative comparisons. (3) We present a novel global color mapping network based on color condition blocks. With about only 35K parameters, it can still achieve state-of-the-art performance. (4) We provide a HDRTV dataset and select five metrics to evaluate SDRTV-to-HDRTV algorithms.

\vspace{-3pt}
\section{Preliminary}
\label{Preliminary}
\textbf{Background.} In this paper, we use SDRTV/HDRTV to represent the content (including image and video) under SDR-TV/HDR-TV standard. The two standards are specified in \cite{rec709, BT1886} and \cite{rec2020, bt2100}, respectively. The basic elements of the HDR-TV standard generally include wide color gamut \cite{rec2020}, PQ or HLG OETF \cite{bt2100} and 10-16 bits depth. In terms of name, ``LDR'' is often used in academia, and ``SDR'' is generally used in industry. Essentially, both of them represent content with low dynamic range but generated from TV production and camera ISP, respectively. For the convenience of distinction and reference, we uniformly use ``LDR-to-HDR'' and ``SDRTV-to-HDRTV'' to represent the conventional image HDR reconstruction and conversion of content from SDR-TV to HDR-TV standard.

\textbf{Explanation.}
Before introducing our method, we first explain that SDRTV-to-HDRTV is functionally different from the previous LDR-to-HDR (i.e., inverse tone mapping) problem. Although the concept of ``HDR'' is involved in these issues, it is undeniable that the connotations of HDR are not the same. It is non-trivial to explain the concepts and differences exhaustively due to the overwhelm of data-level details. In general, the previous LDR-to-HDR methods aim to predict the luminance of images in the linear domain, which is the physical brightness of the scene. In contrast, our goal is to predict HDR images with the HDR \emph{display format} in the pixel domain, which are encoded in HDR-TV standards, such as HDR10, HLG and Dolby Vision. Essentially, content in HDR-TV standard can also be generated from HDR scene radiance. However, the process is an engineering problem and still requires a lot of operations, such as tone mapping and gamut mapping. Therefore, the methods of these two different tasks are not generalizable.

\section{Analysis}
\label{Analysis}
In this section, we first present a simplified SDRTV/ HDRTV formation pipeline which contains the most critical steps in actual production. Then, based on the analysis of the formation pipeline, we propose a novel three-step solution based on the idea of ``divide-and-conquer''. Finally, we compare different solution pipelines of previous methods.

\subsection{SDRTV/HDRTV Formation Pipeline}
To further understand the task of SDRTV-to-HDRTV, we introduce a simplified formation pipeline of SDRTV and HDRTV based on camera ISP and HDR-TV content production \cite{bt2390} as depicted in Fig. \ref{Figure 1 (a)}. Although there are some operations we do not mention here, such as denoising and white balance in camera pipeline and color grading in HDR content production, we retain the four key operations that lead to the difference between SDRTV and HDRTV, which are tone mapping, gamut mapping, opto-electronic transfer function and quantization. In the following equations, we use the subscript ``S'' to represent SDRTV and ``H'' to represent HDRTV. More details about the pipelines can be found in the supplementary material.

\textbf{Tone mapping}. Tone mapping is used to transform the high dynamic range signals to low dynamic range signals for adapting different display devices. The process includes global tone mapping \cite{drago2003adaptive, reinhard2002parameter, tumblin1993tone} and local tone mapping \cite{larson1997visibility, lischinski2006interactive}. Global tone mapping processes all pixels equally with the same function, while parameters are generally related to global image statistics (e.g., average luminance). Local tone mapping can be adaptive to local content and human preference but often brings high computational cost and artifacts. Thus, global tone mapping is mainly used in the HDRTV/SDRTV image formation. The function of global tone mapping can be denoted as:
\begin{equation}
   I_{tS}=T_{S}(I|\theta_{S}),\ I_{tH}=T_{H}(I|\theta_{H}),
\end{equation}
where $T_{S}$ and $T_{H}$ represent the specific tone mapping functions, $ \theta_{S} $ and $ \theta_{H} $ are coefficients related to image statistics. It is noteworthy that S-shape curves are commonly used for global tone mapping and clipping operations often exist in actual process of tone mapping.

\textbf{Gamut mapping}. Gamut mapping is to convert colors from source gamut to target gamut while still preserving the overall look of the scene. According to the standards of ITU-R \cite{rec709} and \cite{rec2020}, the transformations from the original XYZ space to SDRTV (rec.709) and HDRTV (rec.2020) can be represented as :
\begin{equation}
  \begin{bmatrix}
      R_{709}\\ G_{709}\\ B_{709}
      \end{bmatrix}=M_{S}
      \begin{bmatrix}
         X \\ Y \\ Z
      \end{bmatrix}, 
    \begin{bmatrix}
      R_{2020}\\ G_{2020}\\ B_{2020}
      \end{bmatrix}=M_{H}
      \begin{bmatrix}
         X \\ Y \\ Z
      \end{bmatrix},
\end{equation}
where $M_{S}$ and $M_{H}$ are $3\times 3$ constant matrices. 

\textbf{Opto-electronic transfer function}. Opto-electronic transfer function abbreviates as OETF. It is used to convert linear optical signals to non-linear electronic signals in the image formation pipeline. For SDRTV, it approximates a gamma exponential function as $I_{fS}=f_{S}(I)=I^{1/2.2}$. For HDRTV, there are several kinds of OETFs for different standards such as PQ-OETF \cite{PQ} for HDR10 standard and HLG-OETF \cite{bt2100} for HLG standard (HLG stands for Hybrid Log-Gamma). We take the PQ-OETF as an example:
\begin{equation}
  I_{fH}=f_{H}(I)=(\frac{a_{1}+a_{2} I^{b_{1}}}{1+a_{3} I^{b_{1}}})^{b_{2}},
\end{equation}
where $a_{1}, a_{2}, a_{3}, b_{1}, b_{2}$ are constants. 

\textbf{Quantization}. After the above operations, the encoded pixel values are quantized with the function:
\begin{equation}
  I_{q}=Q(I,n)=\frac{\lfloor (2^{n}-1)\times I+0.5 \rfloor}{2^{n}-1},
\end{equation}
where $n$ is 8 for SDRTV and 10-16 for HDRTV.

In summary, the HDRTV and SDRTV content formation pipelines are given by:
\begin{equation}
   I_{S}=Q_{S}\circ f_{S}\circ M_{S}\circ T_{S}(I_{raw}),
\end{equation}
\begin{equation}
   I_{H}=Q_{H}\circ f_{H}\circ M_{H}\circ T_{H}(I_{raw}),
\end{equation}
where $\circ$ represents the connection between two operations. 

\subsection{Proposed Solution Pipeline}
\begin{figure}[!t]
\centering

\subfigure[SDRTV/HDRTV formation pipeline]{
\label{Figure 1 (a)}
\begin{minipage}[t]{\linewidth}
\centering
\vspace{-3pt}
\includegraphics[width=1\linewidth]{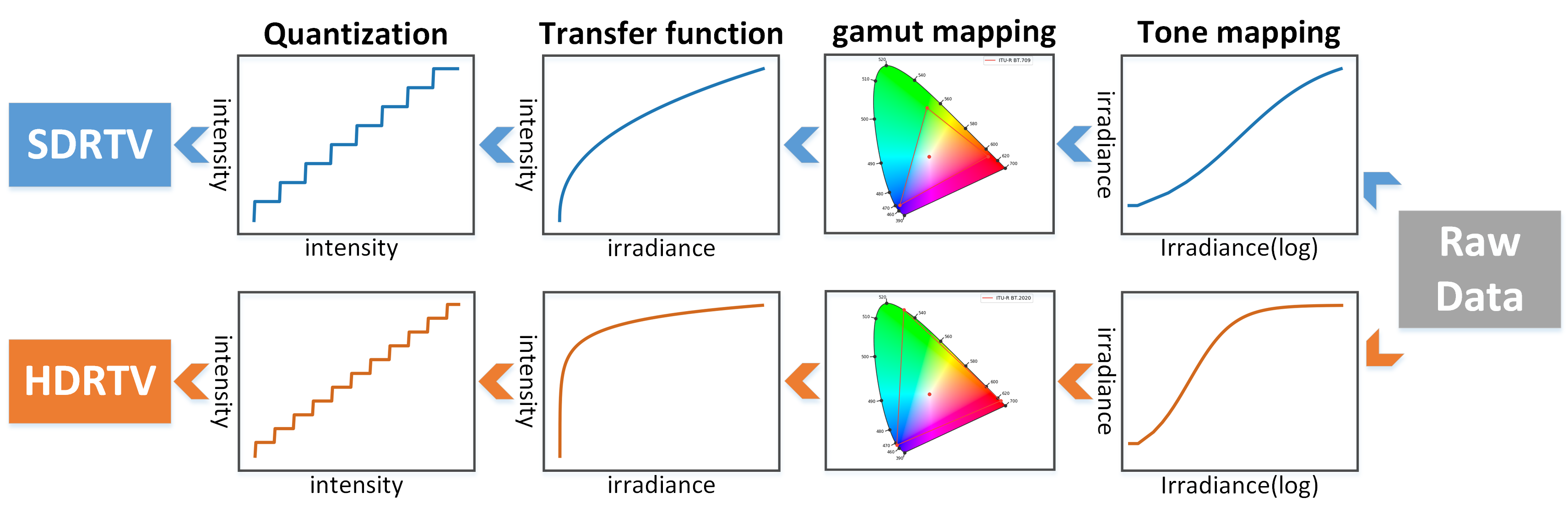}
\end{minipage}%
}
\subfigure[end-to-end solution]{
\label{Figure 1 (c)}
\begin{minipage}[t]{\linewidth}
\centering
\vspace{-3pt}
\includegraphics[width=1\linewidth]{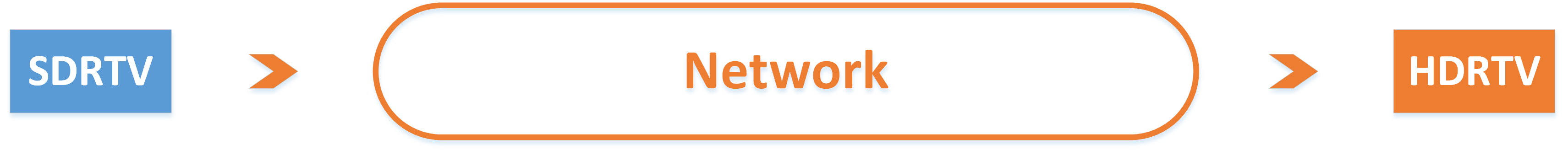}
\end{minipage}%
}
\subfigure[LDR-to-HDR based solution]{
\label{Figure 1 (b)}
\begin{minipage}[t]{\linewidth}
\centering
\vspace{-3pt}
\includegraphics[width=1\linewidth]{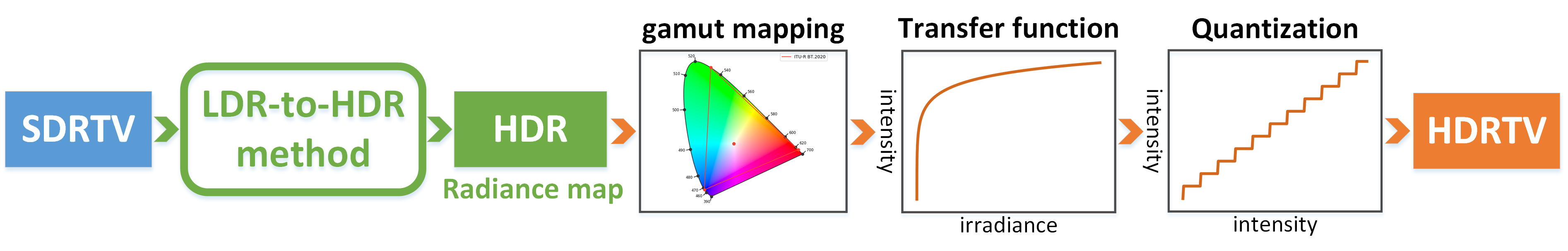}
\end{minipage}%
}
\subfigure[proposed solution pipeline]{
\label{Figure 1 (d)}
\begin{minipage}[t]{\linewidth}
\centering
\vspace{-10pt}
\includegraphics[width=1\linewidth]{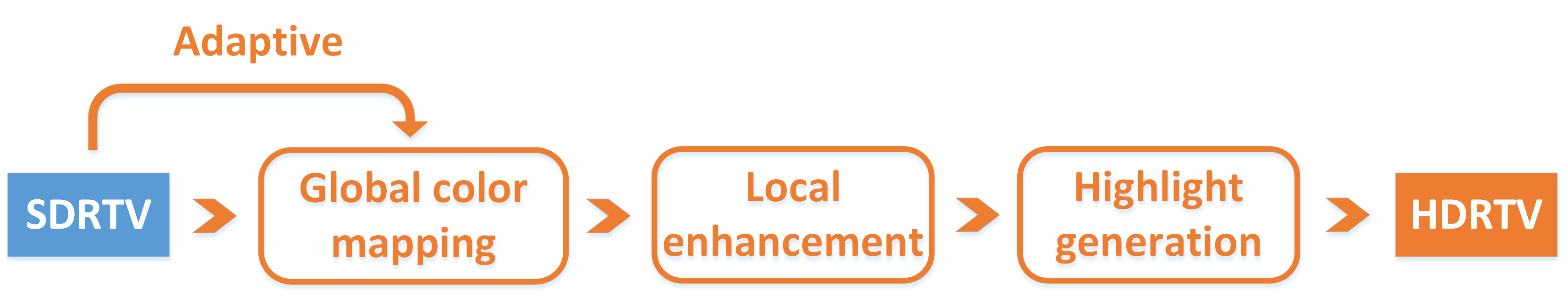}
\end{minipage}%
}
 \caption{Analysis of SDRTV-to-HDRTV. (a) Simplified SDRTV
 /HDRTV content formation pipeline. (b) Existing end-to-end solution pipeline. (c) Solution pipeline based on LDR-to-HDR methods. (d) Proposed solution pipeline. Please refer to supplementary file for detailed explanations of the above diagrams and curves.}
\label{Figure 1 image pipeline}
\vspace{-7pt}
\end{figure}

According to the above formation pipeline, the process of SDRTV-to-HDRTV can be formulated as:
\begin{equation}
   I_{H}=Q_{H}\circ f_{H}\circ M_{H}\circ T_{H} \circ T_{S}^{-1}\circ M_{S}^{-1}\circ f_{S}^{-1}\circ Q_{S}^{-1}(I_{S}),
\end{equation} 
where $T_{S}^{-1}, M_{S}^{-1}, f_{S}^{-1}, Q_{S}^{-1}$ denote the inversions of the corresponding operations. 

We obtain the following observations and insights based on the formation pipeline. Firstly, many critical operations in the formation pipeline are global operations, such as global tone mapping, OETF and gamut mapping. Moreover, the inversions of these operations are also global operations or can be approximately equal to global operations. These operations can be processed using global operators. Secondly, some operations such as local tone mapping and dequantization rely on local spatial information, which can be processed by local operators. Thirdly, there is severe information compression/loss. For example, highlight areas are generally processed through a shoulder operation or simply clipped by tone mapping. 

Inspired by the observations, we propose a new SDRTV-to-HDRTV solution pipeline using the idea of ``divide and conquer''. Our method includes three steps, as shown in Fig. \ref{Figure 1 (d)}. The first step is adaptive global color mapping, which aims at dealing with global operations. This step roughly and adaptively converts the input from SDRTV domain to HDRTV domain. Then, we perform local enhancement, which utilizes local information to enhance the result of the first step. Finally, the highlight generation step is to restore the lost information in the overexposed regions. 
\subsection{Comparison with Existing Solutions}
\label{Comparison with existing solutions}
As an actual industrial problem, SDRTV-to-HDRTV is rarely discussed in academia. In this part, we elaborate on the two groups of existing solutions.

\textbf{End-to-end solution.} Image-to-Image translation methods \cite{isola2017image, zhu2017unpaired} and joint SDRTV-to-HDRTV and super-resolution methods \cite{kim2019deep, kim2020jsi} learn a direct mapping with an end-to-end model to solve the problem, as shown in Fig. \ref{Figure 1 (c)}. However, their methods do not consider the imaging mechanism of SDRTV and HDRTV, thus the results contain some obvious local artifacts and unnatural colors.

\textbf{LDR-to-HDR based solution.} LDR-to-HDR is discussed a lot in academia \cite{rempel2007ldr2hdr, banterle2006inverse, masia2017}. Although these methods are dedicated to predicting HDR scene radiance, it is still necessary to compare them with ours. As in Fig.\ref{Figure 1 (b)}, the HDR radiance map generated by LDR-to-HDR methods needs to go through color gamut mapping to rec.2020, applying PQ or HLG OETF and quantization. For HDR10 standard, the results are obtained by setting the maximum brightness to 1000 $cd/m^2$. Since these steps need to be adjusted according to different data in actual processing, it is hard to get a fair comparison. In this paper, we use the same processing pipeline as \cite{kim2019deep, kim2020jsi} to compare with LDR-to-HDR methods.
\vspace{-4pt}
\section{Method}
\begin{figure*}[!t]
   \begin{center}
   \includegraphics[width=1\linewidth, height=0.32\linewidth]{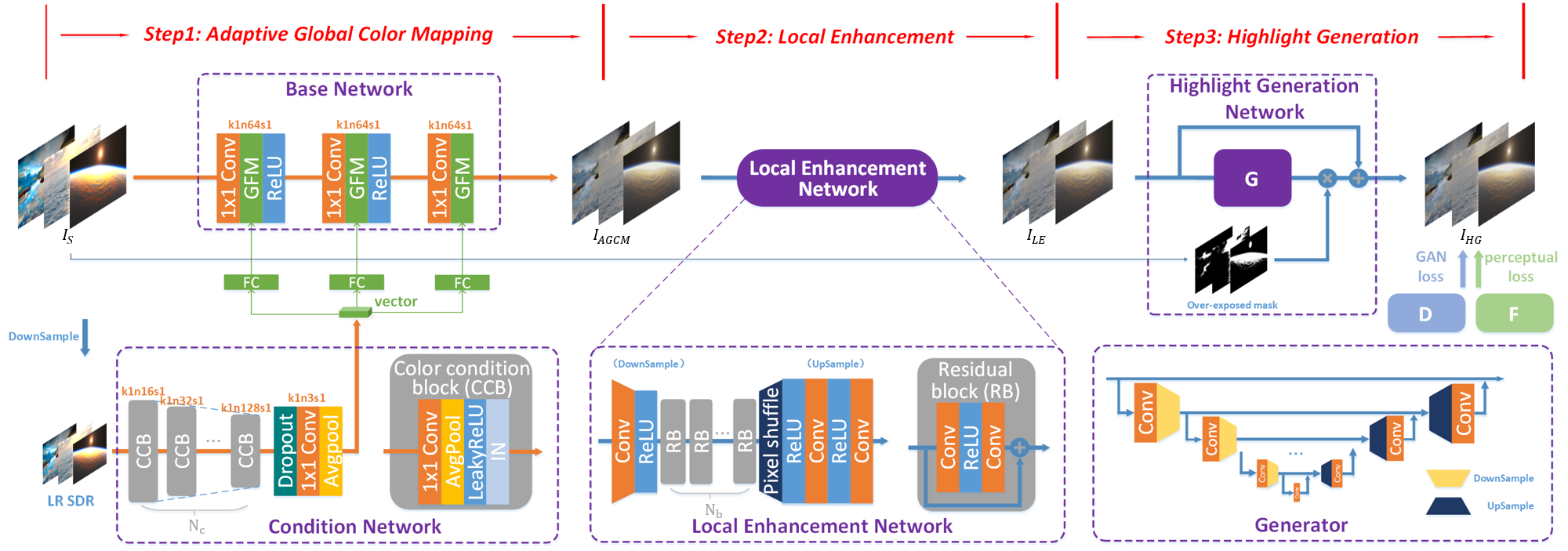}
   \end{center}
   \vspace{-8pt}
      \caption{The architecture of the proposed three-step SDRTV-to-HDRTV method. Each step has a corresponding network.}
     \vspace{-7pt}
   \label{Figure 2 architecture of three-step SDR-to-HDR}
   \end{figure*}

We propose HDRTVNet, a cascaded method consisting of three deep networks for SDRTV-to-HDRTV. Each network corresponds to each step of the solution pipeline.

\vspace{-4pt}
\subsection{Adaptive Global Color Mapping}
As the first step, adaptive global color mapping (AGCM) aims to achieve image-adaptive color mapping from SDR- TV to HDRTV domain. As shown in Fig. \ref{Figure 2 architecture of three-step SDR-to-HDR}, the proposed model consists of a base network and a condition network.

\subsubsection{Base Network}
The base network is designed to handle global operations, which works on each pixel independently. This global mapping can be denoted as:
\begin{equation}
   I_{B}(x,y)=f(I_{S}(x,y)), \forall (x,y)\in I_{S},
\end{equation}
where $(x,y)$ represents the coordinate of the pixel in the image $I$ and $I_{B}$ represents the output of base network. As demonstrated in CSRNet \cite{liu2021very}, a fully convolution network with only $1\times 1$ convolutions and activation functions can achieve this kind of global mapping. Thus, our base network is composed of $N_l$ convolutional layers with $1\times 1$ filters and $N_l$-1 ReLU activations, which can be denoted as:
\begin{equation}
   I_{B}=Conv_{1\times 1}\circ (ReLU\circ Conv_{1\times 1})^{N_l-1}(I_{S}).
\end{equation}
The proposed base network takes an 8-bits SDRTV image as input and generates an HDRTV image encoded with 10-16 bits. Although the base network can only learn a one-to-one color mapping, it also achieves considerable performances, as shown in Tab. \ref{Table 1 comparison}. It is noteworthy that the base network can perform like a 3D lookup table (3D LUT) with fewer parameters rather than learning 3D LUT directly, and please refer to supplementary material for more results.

\subsubsection{Condition Network}
The global priors are indispensable for adaptive global color mapping. For example, global tone mapping requires global image statistics. To achieve image-adaptive mapping, we add a condition network to modulate the base network. Previous works \cite{wang2018recovering, he2020conditional, chen2021hdrunet} usually adopt the prior of image content, where the condition network can extract spatial and local information from the input image by $N_k\times N_k$ ($N_k>1$) filters. However, for SDRTV-to-HDRTV problem, we find that global mapping is conditioned on global image statistics or color distribution. This kind of color condition is independent of spatial information, thus it is inappropriate to adopt previous structures of condition in this problem. Our proposed condition network focuses on extracting information related to color to realize adjustable mapping. As shown in Fig. \ref{Figure 2 architecture of three-step SDR-to-HDR}, the proposed condition network consists of several color condition blocks (CCB), convolution layers, feature dropout and global average pooling.

\textbf{Color condition block}. A color condition block contains a convolution layer with $1\times 1$ filters, an average pooling layer, a LeakyReLU activation and an instance normalization layer \cite{dumoulin2016learned}, which can be written as follows:
\begin{equation}
   CCB(\cdot)=IN\circ LReLU\circ avgpool\circ Conv_{1\times 1}(\cdot),
\end{equation}
where $\cdot $ denotes the input of CCB. The condition network takes a down-sampled SDRTV image as input and outputs a condition vector $V$. Our condition network is denoted by:
\begin{equation}
   V=GAP\circ Conv_{1\times 1}\circ Dropout\circ CCB^{N_c}(I_{S}).
\end{equation}
Since the convolutional layers only contain $1\times 1$ filters, the condition network can not extract local features. With the help of pooling layers, the network can extract global priors based on image statistics. To avoid overfitting, we add a dropout before the last convolutional layer and global average pooling. It performs like adding a multiplicative Bernoulli noise on features which has an effect similar to data augmentation. It is worth noting that even if we take the image by shuffling the pixels randomly as input, we can also obtain comparable performance with the correct arrangement of pixels. It suggests that the effective prior is not related to spatial information in global color mapping.

\subsubsection{Global Feature Modulation}
To utilize the extracted global priors, we introduce the global feature modulation (GFM) \cite{he2020conditional} strategy to modulate the base network, which has been successfully applied in photo retouching tasks. Through GFM, the intermediate features of the base network can be modulated by scaling and shifting operations. It can be described as:
\begin{equation}
   GFM(x_{i})=\alpha _{1}*x_{i}+\alpha _{2},
\end{equation}
where $x_{i}$ denotes the feature map. $\alpha _{1}$ and $\alpha _{1}$ represent the scale and shift factor, respectively.

Overall, AGCM network can be formulated as:
\begin{equation}
    \begin{split}
   I_{AGCM}=GFM\circ Conv_{1\times 1}\circ(ReLU\circ \\GFM \circ Conv_{1\times 1})^{N_l-1}(I_{S}),
   \end{split}
\end{equation}
where $I_{AGCM}$ denotes the output of adaptive global color mapping. To optimize adaptive global color mapping, we minimize the distance of the output and the ground truth HDRTV image using L2 loss function.

\subsection{Local Enhancement}
Local enhancement (LE) is performed followed by AGCM. Although AGCM can obtain considerable performance, local enhancement is indispensable for SDRTV-to-HDRTV. It is a remarkable fact that if we directly use local operations to learn end-to-end mapping before adaptive global color mapping, the output results often have obvious artifacts. Details can be founded in the supplementary material.

To achieve local enhancement, we directly adopt a classical ResNet structure \cite{he2016identity}. More advanced architectures can be used here, but this is not the focus of this work. Details of the enhancement network are described as follows. This step takes the output of adaptive color mapping as input. During this process, the input is down-sampled by the first convolutional layer, then goes through several residual blocks (RB), and finally is up-sampled to the original size as output. The overall operation can be represented as:
\begin{equation}
    \begin{split}
   I_{LE}=Conv\circ ReLU\circ Conv\circ ReLU\circ \\PS\circ RB^{N_b}\circ ReLU\circ Conv(I_{AGCM}),
    \end{split}
\end{equation}
where $M$ is the number of residual blocks. $L_{LE}$ is the HDR image generated by the proposed local enhancement network. L2 loss function is adopted for local enhancement.

\subsection{Highlight Generation}
The third step of our solution pipeline is highlight generation (HG), which aims at hallucinating some details that were lost due to extreme compression. To achieve this goal, we adopt a generative adversarial network (GAN) \cite{goodfellow2014generative} that has the potential to generate details. This network consists of a generator and a discriminator. We adopt an encoder-decoder architecture with skip connections \cite{ronneberger2015u} as generator and an commonly used VGG architecture \cite{simonyan2014very} as discriminator. The formulation of this step can be represented as:
\begin{equation}
   I_{o}=I_{mask}\odot G(I_{i})+I_{i},
\end{equation}
where $G$ denotes the generator and $\odot $ denotes elementwise multiplication. The mask $I_{mask}$ can be computed by $p=max(I_{i}-\gamma,0)/1-\gamma$ as \cite{liu2020single}. For learning highlight generation network, we joint optimize three types of losses, including L1 loss, perceptual loss and GAN loss, which can be formulated as:
\begin{equation}
   L_{HG}=\alpha L_{1}+\beta L_{P}+\gamma L_{GAN},
\end{equation}
where $\alpha, \beta, \gamma$ are loss weights. The perceptual loss is to measures the similarity in feature space, as $L_{P}=\Vert \psi (I_{HG})-\psi (I_{LE}) \Vert_{2}^{2}$, where $\psi{(I)}$ represents the feature maps of image $I$. The GAN loss is denoted as $L_{G}=-\log D(G(I_{LE}))$, where $D$ denotes the discriminator.

\section{Experiments}
\label{Experiments}
\subsection{Experimental Setup.}
\textbf{Dataset.} We collect 22 videos under HDR10 standard and their counterpart SDR version from YouTube as \cite{kim2019deep}. All of these HDR videos are encoded by PQ-OETF and rec.2020 gamut. 18 video pairs are used for training and the left for testing. To avoid high coherence among extracted frames, we sample a frame every two seconds of each video and generate a training set with 1235 images. 
Besides, 117 images without repeated scenes are selected as the test set.

\textbf{Training details.} For the proposed AGCM, the base network consists of 3 convolution layers with $1\times 1$ kernel size and 64 channels, and the condition network contains 4 CCBs. For LE, all convolution filters are of size $3\times 3$ with 64 output channels, except for the convolution layer in the pixel shuffle module\cite{shi2016real} with 256 channels, and the output layer with 3 channels. The number of the RBs is 16. Strides of all layers are set to 1 except for the first convolution layer with a stride of 2. For the part of HG, there are five convolution layers followed by max-pooling operation and also five convolution layers with pixel shuffle. Each filter has kernel size of $3\times 3$. The number of channels is increased from 64 to 1024 in the downsampling process and reverses in the process of upsampling. More implementation details can be found in the supplementary material. 

\begin{table*}[t]
\centering
\resizebox{160mm}{35mm}{%
\begin{tabular}{cc|c|ccccc}
\hline
\multicolumn{2}{c|}{Method} &Params$\downarrow$ & PSNR$\uparrow$ & SSIM$\uparrow$ & SR-SIM$\uparrow$ & $\Delta E_{ITP}\downarrow$ & HDR-VDP3$\uparrow$ \\ \hline
 & HuoPhyEO \cite{huo2014physiological} & - & 25.90 & 0.9296 & 0.9881 & 38.06 & 7.893 \\ \cline{2-8} 
\multirow{-2}{*}{LDR-to-HDR} & KovaleskiEO \cite{kovaleski2014high} & - & 27.89 & 0.9273 & 0.9809 & 28.00 & 7.431 \\ \hline
 & ResNet \cite{he2016identity} & 1.37M & {\color[HTML]{3166FF} \textbf{37.32}} & 0.9720 & 0.9950 & {\color[HTML]{3166FF} \textbf{9.02}} & 8.391 \\ \cline{2-8} 
 & Pixel2Pixel \cite{isola2017image} & 11.38M & 25.80 & 0.8777 & 0.9871 & 44.25 & 7.136 \\ \cline{2-8} 
\multirow{-3}{*}{\begin{tabular}[c]{@{}c@{}}image-to-image\\ traslation\end{tabular}} & CycleGAN \cite{zhu2017unpaired} & 11.38M & 21.33 & 0.8496 & 0.9595 & 77.74 & 6.941 \\ \hline
 & HDRNet \cite{gharbi2017deep} & 482K & 35.73 & 0.9664 & 0.9957 & 11.52 & 8.462 \\ \cline{2-8} 
 & CSRNet \cite{he2020conditional} & {\color[HTML]{009901} \textbf{36K}} & 35.04 & 0.9625 & 0.9955 & 14.28 & 8.400 \\ \cline{2-8} 
\multirow{-3}{*}{\begin{tabular}[c]{@{}c@{}}photo\\ retouching\end{tabular}} & Ada-3DLUT \cite{zeng2020learning} & 594K & 36.22 & 0.9658 & {\color[HTML]{009901} \textbf{0.9967}} & 10.89 & 8.423 \\ \hline
 & Deep SR-ITM \cite{kim2019deep} & 2.87M & 37.10 & 0.9686 & 0.9950 & 9.24 & 8.233 \\ \cline{2-8} 
\multirow{-2}{*}{SDRTV-to-HDRTV} & JSI-GAN \cite{kim2020jsi} & 1.06M & 37.01 & {\color[HTML]{009901} \textbf{0.9694}} & 0.9928 & 9.36 & 8.169 \\ \hline
 & Base Network & {\color[HTML]{FE0000} \textbf{5K}} & 36.14 & 0.9643 & 0.9961 & 10.43 & 8.305 \\ \cline{2-8} 
 & AGCM & {\color[HTML]{3166FF} \textbf{35K}} & 36.88 & 0.9655 & 0.9964 & 9.78 & {\color[HTML]{009901} \textbf{8.464}} \\ \cline{2-8} 
 & AGCM+LE & 1.41M & {\color[HTML]{FE0000} \textbf{37.61}} & {\color[HTML]{FE0000} \textbf{0.9726}} & {\color[HTML]{3166FF} \textbf{0.9967}} & {\color[HTML]{FE0000} \textbf{8.89}} & {\color[HTML]{FE0000} \textbf{8.613}} \\ \cline{2-8} 
\multirow{-4}{*}{\begin{tabular}[c]{@{}c@{}}HDRTVNet\\ (ours)\end{tabular}} & AGCM+LE+HG & 37.20M & {\color[HTML]{009901} \textbf{37.21}} & {\color[HTML]{3166FF} \textbf{0.9699}} & {\color[HTML]{FE0000} \textbf{0.9968}} & {\color[HTML]{009901} \textbf{9.11}} & {\color[HTML]{3166FF} \textbf{8.569}} \\ \hline
\end{tabular}%
}
\vspace{2pt}
\caption{Quantitative comparisons. {\color[HTML]{FE0000} Red} text indicates the best, {\color[HTML]{3166FF} blue} text indicates the second and {\color[HTML]{009901} green} text indicates the third.}
\label{Table 1 comparison}
\end{table*}

\subsection{Evaluation of SDRTV-to-HDRTV}
\textbf{Evaluation metrics.} We employ five evaluation metrics for comprehensive comparisons, including PSNR, SSIM, SR-SIM \cite{zhang2012sr}, HDR-VDP3 \cite{mantiuk2011hdr} and $\Delta E_{ITP}$ \cite{ITP}. SSIM and SR-SIM are commonly used to measure image similarity. Although they are designed to evaluate SDR image, \cite{athar2019perceptual} shows that SR-SIM has a favorable performance of evaluation for HDR standard. Besides, we introduce $\Delta E_{ITP}$ to measure the color difference, which is designed for HDRTV. HDR-VDP3 is a new improved version of HDR-VDP2 that supports the rec.2020 gamut. To employ HDR-VDP3, results are compared by setting ``side-by-side'' task, ``rgb-bt.2020'' color encoding, 50 pixel per degree and option of ``rgb-display'' with ``led-lcd-wcg''.

\textbf{Visualization.} We directly show the HDRTV images encoded in 16-bits ``PNG'' format without extra processing. Since HDRTV images are decoded by gamma EOTF on SDR screens, they may look relatively darker than on HDR screens. Previous work \cite{kim2019deep, kim2020jsi} shows HDRTV images by software for visualization. However, it introduces an unfair comparison since the video player may reduce the unnatural artifacts of original HDRTV images. In contrast, our visualization method preserves the details in highlight areas and compares all methods in the same conditions. Tone mapped results can be founded in the supplementary file.

\subsection{Comparison with Other Methods}

\begin{figure*}[!t]
   \begin{center}
   \includegraphics[scale=0.3]{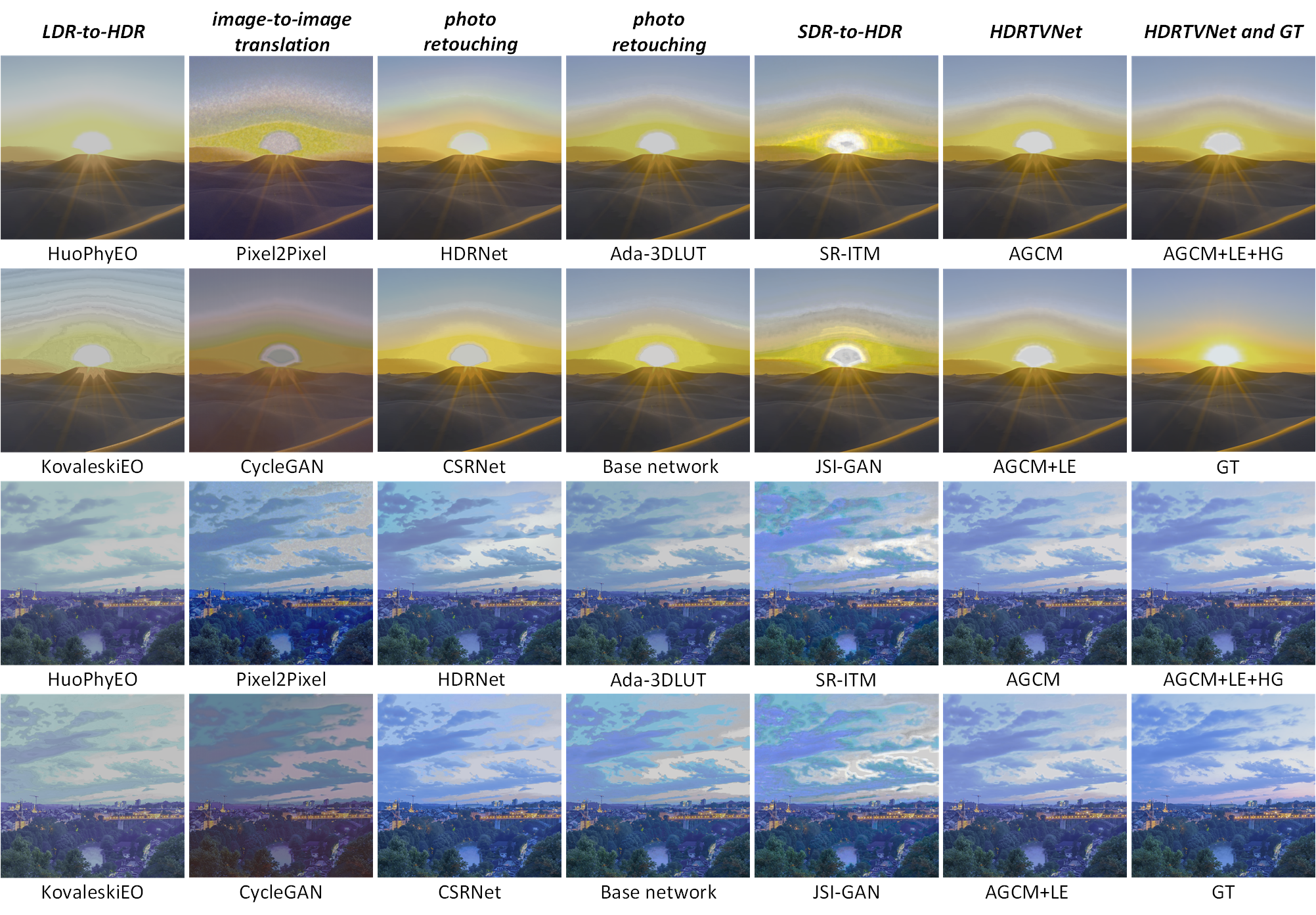}
   \end{center}
   \vspace{-10pt}
      \caption{Qualitative comparisons. The top row describes the categories of algorithms.}
   \label{Figure 3 visualization}
   \end{figure*}

\textbf{Compared methods.} 
We compare our results with four types of methods including joint SR with SDRTV-to-HDRTV, image-to-image translation, photo retouching and LDR-to-HDR. Since these methods are not completely designed for this task, we have done the necessary processing of these methods. For joint SR with SDRTV-to-HDRTV methods, we change the stride of the first convolutional layer to 2 for downsampling to match the size of input and output\footnote{We have also conducted experiments to remove the pixel shuffle layer instead of downsampling at the beginning, but the results show that it could not bring improvements but increase the computational cost significantly.}. For LDR-to-HDR methods, we process the results as illustrated in Sec. \ref{Comparison with existing solutions}. Note that we adopt the same processing steps as \cite{kim2019deep, kim2020jsi}. All data-driven methods are retrained on the proposed dataset. 

\textbf{Quantitative comparison.} As shown in Tab. \ref{Table 1 comparison}, our method HDRTVNet outperforms other methods by a large margin on all evaluation metrics. It is worth noticing that our first step AGCM could already achieve comparable performance to Ada-3DLUT, but with only 1/17 of its parameters. The LDR-to-HDR based solutions have poor results as their pipeline is different from ours. It is hard to compare with them on a completely fair platform.

\textbf{Qualitative comparison.} The results of qualitative comparisons are shown in Fig.\ref{Figure 3 visualization}. LDR-to-HDR based methods and image-to-image translation methods tend to generate low-contrast images. All approaches of LDR-to-HDR based, image-to-image translation and SDRTV-to-HDRTV produce unnatural colors and obvious artifacts except for HuoPhyEO \cite{huo2014physiological}. Outputs generated by photo retouching methods are relatively better but suffer from the color cast. Our method HDRTVNet could produce natural colors and high contrast as referred ground truth and do not introduce any artifacts. Further, the visual quality improves with more processing steps, i.e., AGCM$<$AGCM+LE$<$AGCM+LE+HG. More results can be found in the supplementary material.

\subsection{Color Transition Test}
\begin{figure*}[!t]
   \begin{center}
   \includegraphics[scale=0.3]{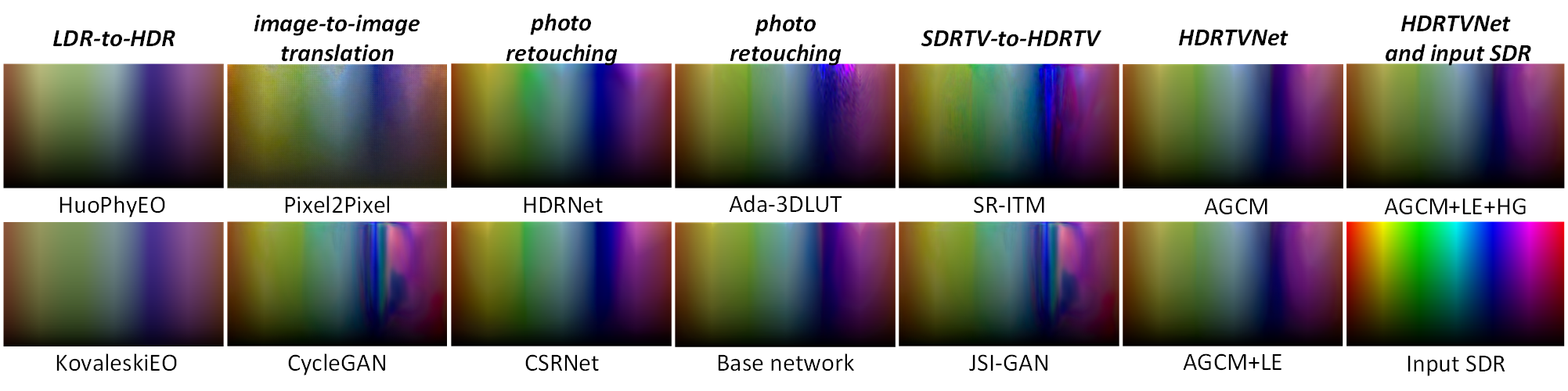}
   \end{center}
      \vspace{-10pt}
      \caption{Color test visualization. The top row describes the categories of algorithms.}
   \label{Figure 4 color transition}
   \end{figure*}
We observe that many previous methods perform poorly in the highlight regions, especially where the color changes. To reveal this phenomenon, we conduct a color transition test using a man-made color card as input image shown in Fig. \ref{Figure 4 color transition}. We obtain the following three observations. First, the unnatural transition and color blending problem can be easily observed in the outputs of several methods, which rely on learning local information (e.g., Deep SR-ITM, JSI-GAN, Pixel2Pixel, CycleGAN) or based on local conditions (e.g., 3D-LUT). Second, our method performs smooth transition even learning local information (e.g., AGCM+LE, AGCM+LE+HG), which shows the superiority of our cascaded solution pipeline. Third, blue regions suffer the most server unnatural color transition. A reasonable explanation is that blue colors are harder to recover than other colors, since more information is missing in the blue area in the process of extreme compression. 

\subsection{Ablation Study}
\begin{figure}[!t]
   \begin{center}
   \includegraphics[scale=0.23]{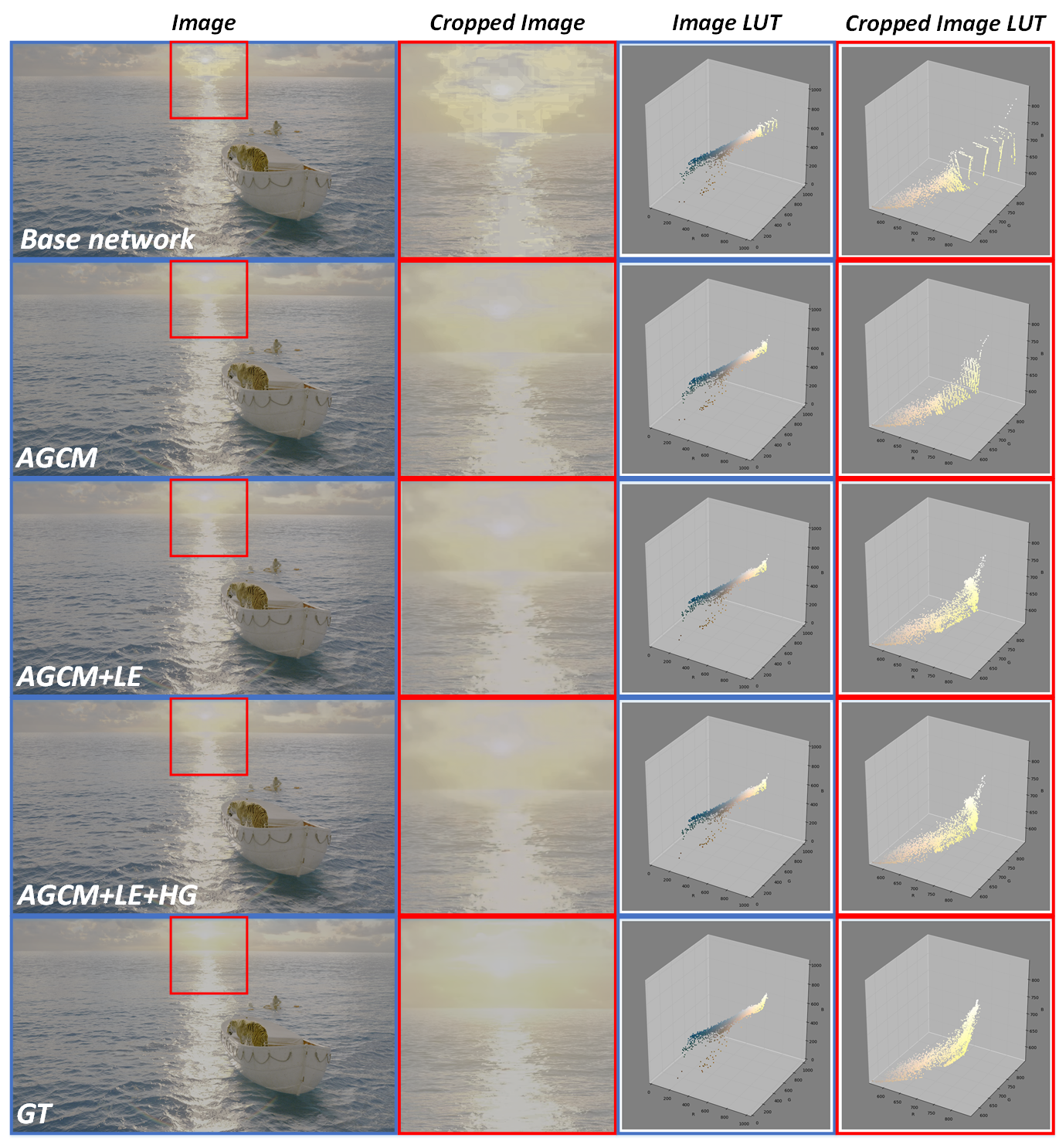}
   \end{center}
    \vspace{-10pt}
      \caption{Visualization of LUTs. The LUT cube shows the position of SDR colors in the HDRTV domain at the coordinates determined by the pixel values of their corresponding HDRTV pixels.}
   \label{Figure 5 LUTs}
   \vspace{-10pt}
\end{figure}

\textbf{Adaptive global color mapping.} The process of adaptive color mapping can be observed by LUT cubes shown in Fig. \ref{Figure 5 LUTs}. The color of each point in the cube corresponds to the input SDRTV color, and its coordinates in the cube correspond to values of HDRTV pixels after the current mapping. Note that if an SDRTV color is mapped to multiple HDRTV colors, the color will also appear in multiple positions of the cube. Since the basic network can only learn a one-to-one color mapping throughout the dataset, the color transition of the LUT manifold in the highlight areas is not smooth. It can be easily seen that the output of the base network suffers severe posterization artifacts. In contrast, the color condition network helps the base network learn image-adaptive color mapping. Then the artifacts disappear, and the LUT becomes concentrated. In Fig. \ref{Figure 5 LUTs}, our method also performs better in the color transition test.

\textbf{Local enhancement.} This part takes the adaptive-global-color-mapped image as input, aiming to handle the one-to-many color mapping and some local operations in the SDRTV-to-HDRTV process. The same color in SDRTV domain is mapped to multiple HDRTV colors, and the color distribution becomes more compact and smooth as in Fig. \ref{Figure 5 LUTs}. Noting that adding local enhancement after adaptive global color mapping achieves the best performance in quantitative comparison (in Tab. \ref{Table 1 comparison}) without introducing artifacts. If we apply the local enhancement as the first step, it will generate apparent local artifacts, similar to other end-to-end mapping methods. Please refer to the supplementary material.

\textbf{Highlight generation.} Highlight generation aims to hallucinate details in the over-exposed regions. We declare that this part is not designed for numerical improvement, but functional increase. Due to the inevitable loss of information in the production (e.g., dynamic range compression), we believe that it is necessary to use some generative methods to deal with these parts. Although our HG method has certain limitations for numerical evaluation, we can still observe that it makes colors in the LUT cube denser and makes the highlight regions look more natural. 

\subsection{User Study}
We conduct a user study with 20 participants for subjective evaluation. Four methods with the best performance in each category are selected to compare with our method and ground truth. Participants are asked to rank them according to the visual quality. 25 images are randomly selected from the testing set and shown to participants on HDR-TV (Sony X9500G with a peak brightness of 1300 nits) in darkroom. More details of how we conduct experiment can be founded in the supplementary material. As suggested in Fig. \ref{Figure 6 Ranking}, our method achieves a better visual ranking than other competitors except for the ground truth. 

\begin{figure}[!h]
    \centering
    \includegraphics[width=8cm]{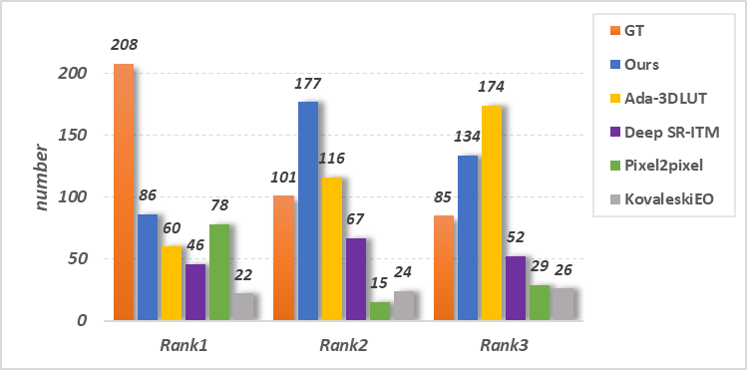}
    \caption{Ranking results of user study. Rank1 means the best subjective feeling.}
    \label{Figure 6 Ranking}
\end{figure}
\vspace{-10pt}

\section{Conclusion}
We have provided a novel SDRTV-to-HDRTV solution pipeline based on the HDRTV/SDRTV formation pipeline using divide-and-conquer. Moreover, we have introduced a new method, HDRTVNet, for the problem. According to the three types of operations in HDRTV/SDRTV formation pipeline including global operation, local operation and shoulder operation, the whole method is divided into adaptive global color mapping, local enhancement and highlight generation correspondingly. Besides, a novel color condition network has been proposed with fewer parameters and better performance than other photo retouching approaches. Comprehensive experiments show the superiority of our solution in quantitative comparison and visual quality. 

\noindent\textbf{Acknowledgement.} This work is partially supported by National Natural Science Foundation of China (61906184), the Science and Technology Service Network Initiative of Chinese Academy of Sciences (KFJ-STS-QYZX-092), the Shanghai Committee of Science and Technology, China (Grant No. 21DZ1100100).

{\small
\bibliographystyle{ieee_fullname}
\bibliography{egbib}
}

\clearpage

\includepdf[pages={1}]{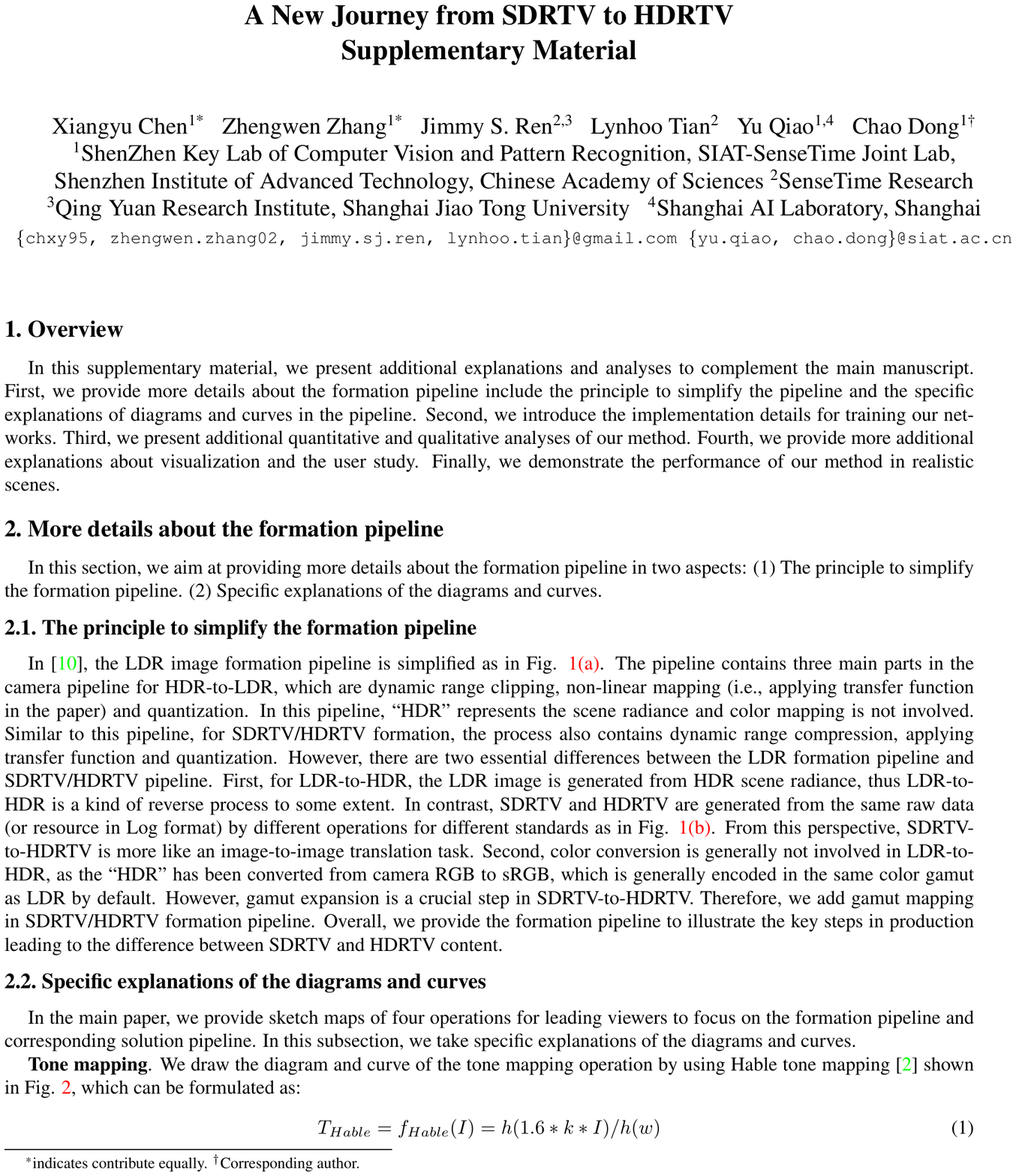}
\includepdf[pages={2}]{supp.pdf}
\includepdf[pages={3}]{supp.pdf}
\includepdf[pages={4}]{supp.pdf}
\includepdf[pages={5}]{supp.pdf}
\includepdf[pages={6}]{supp.pdf}
\includepdf[pages={7}]{supp.pdf}
\includepdf[pages={8}]{supp.pdf}
\includepdf[pages={9}]{supp.pdf}

\end{document}